\documentclass[pdftex,twocolumn,epjc3upd,final]{svjour3}  
\smartqed  
\RequirePackage{graphicx}
\RequirePackage{amsmath}
\RequirePackage{esvect}
\RequirePackage{lineno}
\RequirePackage[T1]{fontenc}
\RequirePackage{xcolor}
\RequirePackage{fix-cm}
\journalname{Eur. Phys. J. A}

\begin{document}

%
%

\title{An experimental program with high duty-cycle polarized and unpolarized positron beams at Jefferson Lab}

\titlerunning{e$^+$@JLab}   

\author{
A.~Accardi\thanksref{add1,add39} 
\and
A.~Afanasev\thanksref{add3} 
\and
I. Albayrak\thanksref{add41} 
\and
S.F.~Ali\thanksref{add56} 
\and
M.~Amaryan\thanksref{add17} 
\and
J.R.M.~Annand\thanksref{add37} 
\and
J.~Arrington\thanksref{add11j}
\and
A.~Asaturyan\thanksref{add59}
\and
H.~Atac\thanksref{add47}
\and
H.~Avakian\thanksref{add1}
% ...
}

\thankstext{Cp}{Contact person: voutier@ijclab.in2p3.fr}

\institute{
% ...
Thomas Jefferson National Accelerator Facility, Newport News, VA 23606, USA \label{add1}
% ...
\and
The George Washington University, Washington, DC 20052, USA \label{add3}
%
%
% ...
\and
Lawrence Berkeley National Laboratory, Berkeley, CA 94720, USA \label{add11j}
% ...
\and
University of Glasgow, Glasgow G12 8QQ, United Kingdom \label{add37}
% ...
\and
Hampton University, Hampton, VA 23668, USA \label{add39}
% ...
\and
Akdeniz \"Universitesi, 07070 Konyaalti/Antalya, Turkey \label{add41}
% ...
\and
Old Dominion University, Norfolk, VA 23529, USA \label{add17}
% ...
\and
Temple University, Physics Department, Philadelphia, PA 19122-180, USA \label{add47}
% ...
\and
The Catholic University of America, Washington, DC 20064, USA \label{add56}
% ...
\and
A.~Alikhanyan National Laboratory, Yerevan Physics Institute, Yerevan 375036, Armenia \label{add59}
%
%
% ...
}

\authorrunning{A.~Accardi {\it et al.}} 

\date{Draft : \today}

\maketitle

%
%

\begin{abstract}

Positron beams, both polarized and unpolarized, are identified as essential ingredients for the experimental programs at the next generation of lepton accelerators. In the context of the hadronic physics program at Jefferson Lab (JLab), positron beams are complementary, even essential, tools for a precise understanding of the electromagnetic structure of nucleons and nuclei, in both the elastic and deep-inelastic regimes. For instance, elastic scattering of polarized and unpolarized electrons and positrons from the nucleon enables a mo\-del independent determination of its electromagnetic form factors. Also, the deeply-virtual scattering of polarized and unpolarized electrons and positrons allows unambiguous separation of the different contributions to the cross section of the lepto-pro\-duction of photons and of lepton-pairs, enabling an accurate determination of the nucleons and nuclei generalized parton distributions, and providing an access to the gravitational form factors.  Furthermore, positron beams offer the possibility of alternative tests of the Standard Model of particle physics through the search of a dark photon, the precise measurement of electro\-weak couplings, and the investigation of charged lepton flavor violation. This document discusses the perspectives of an  experimental program with high duty-cycle positron beams at JLab.


\end{abstract}

%

\section{Introduction}

Quantum Electrodynamics (QED) is an outstanding example of the power of quantum theory. The highly accurate predictive power of this theory allows us not only to investigate numerous physics phenomena at the macroscopic, atomic, nuclear, and partonic scales, but also to test the validity of the Standard Model of particle physics. Therefore, QED promotes electrons and positrons as unique phy\-sics probes, as demonstrated worldwide over decades of scientific research at different laboratories.

Both from the projectile and the target points of view, spin appears nowadays as the finest tool for the study of the inner structure of matter. Recent examples from the experimental physics program developed at the Thomas Jefferson National Accelerator Facility (JLab) include: the measurement of polarization observables in elastic electron scattering off the nucleon~\cite{Jones:1999rz,Gayou:2001qd,Puckett:2010ac}, which established the unexpected magnitude and behavior of the proton electric form factor at high momentum transfer (see \cite{Punjabi:2015bba} for a review); the experimental evidence, in the production of real photons from a polarized electron beam interacting with unpolarized protons, of a strong sensitivity to the electron beam helicity~\cite{Stepanyan:2001sm}, that opened the investigation of the 3-dimensional partonic structure of nucleons and nuclei via the generalized parton distributions (GPDs)~\cite{Mueller:1998fv} measured through the deeply-virtual Compton scattering (DVCS)~\cite{Ji:1996ek,Radyushkin:1997ki}; the achievement of a unique parity violation experimental  program~\cite{Armstrong:2005hs,Aniol:2005zf,Aniol:2005zg,Acha:2006my,Androic:2009aa,Androic:2011rh,G0:2011aa,Androic:2013rhu,Androic:2018kni} that accessed the smallest polarized beam asymmetries ever measured (a few $\times 10^{-7}$) and provided the first determination of the weak charge of the proton~\cite{Androic:2018kni}, along with the first nonzero observation of the neutral current electron-quark vector-axial coupling~\cite{Wang:2014bba}, allowing for stringent tests of the Standard Model at the TeV mass scale~\cite{Young:2006jc}; {\it etc}. Undoubtedly, polarization became an important capability and a mandatory property of the current and next generation of accelerators. 

The combination of the QED predictive power and the fineness of the spin probe led to a large but yet limited variety of  impressive physics results. Adding to this tool-kit charge symmetry properties in terms of polarized positron beams will provide a more complete and accurate picture of the physics at play, whatever physics scale is involved~\cite{Voutier:2014kea}. In the context of the experimental study of the structure of hadronic matter carried out at JLab, the electromagnetic interaction dominates lepton-hadron reactions and there is no intrinsic difference between the physics information obtained from the scattering of electrons or positrons off an hadronic target. However, when a reaction process is a combination of more than one elementary QED-mechanism, the comparison between electron and posi\-tron scattering allows us to isolate their quantum interference. This is of particular interest for studying limitations of the one-photon exchange Born approximation in elastic and inelastic scatterings~\cite{Guichon:2003qm,Blunden:2003sp}. It is also essential for the experimental determination of the GPDs where the interference between the known Bethe-Heitler (BH) process and the unknown DVCS requires polarized and unpolarized electron and positron beams for a model independent extraction of the different contributions to the cross section~\cite{Voutier:2014kea}. Such polarized lepton beams also provide the ability to test new physics beyond the frontiers of the Standard Model via a precise measurement of the electroweak coupling parameters~\cite{Zheng:2021hcf}, the investigation of charged lepton flavor violation~\cite{YulSon}, and the search for new particles linked to dark matter~\cite{Wojtsekhowski:2009vz,Marsicano:2018oqf}.

The production of high-quality polarized positron beams to suit these many applications remains however a highly difficult  task that, until recently, was feasible only at large scale accelerator facilities. Relying on the most recent advances in high polarization and high intensity electron sources~\cite{Adderley:2010zz}, the PEPPo (Polarized Electrons for Polarized Positrons) technique~\cite{Abbott:2016hyf}, demonstrated at the injector of the Continuous Electron Beam Accelerator Facility (CEBAF) of JLab, provides a novel and widely accessible approach based on the production, within a high-$Z$ target, of polarized $e^+e^-$ pairs from the circularly polarized bremsstrah\-lung radiation of a low energy highly polarized electron beam~\cite{Olsen:1959zz,Kuraev:2009uy}. As opposed to other schemes operating at GeV lepton beam energies~\cite{Sokolov:1963zn,Omori:2005eb,Alexander:2008zza}, the operation of the PEPPo technique requires only energies above the pair-produc\-tion threshold and is thus ideally suited for a polarized positron beam at CEBAF.

This document aims at an introduction to the Topical Issue of the European Physics Journal A discussing the physics case of {\it Positron beams and physics at Jefferson Lab (e$^+$@JLab)}. It presents the main physics merits of an experimental program with high energy positron beams at JLab. The next sections discuss their benefits for the investigation of two-photon exchange mechanisms, for the study of the partonic structure of nucleons and nuclei, and for testing the Standard Model. The last section addresses the production and implementation of polarized and unpolarized positron beams at JLab. 

%
%

\section{Two-photon exchange physics}

Measuring the differences between positron scattering and electron scattering is one of the best ways to isolate the effects of two-photon exchange (TPE). 
The leading contribution of TPE beyond the one-photon exchange level (OPE) is the interference between OPE and TPE, which changes sign with a reversal of lepton charge. A positron source at CEBAF would open the possibility of constraining TPE through a number of observables, some of which have never been measured before (see~\cite{Afanasev:2017gsk} for a recent review of the status of TPE in elastic electron-proton scattering). 

\begin{figure}[t!]
    \centering
    \includegraphics[width=0.99\columnwidth]{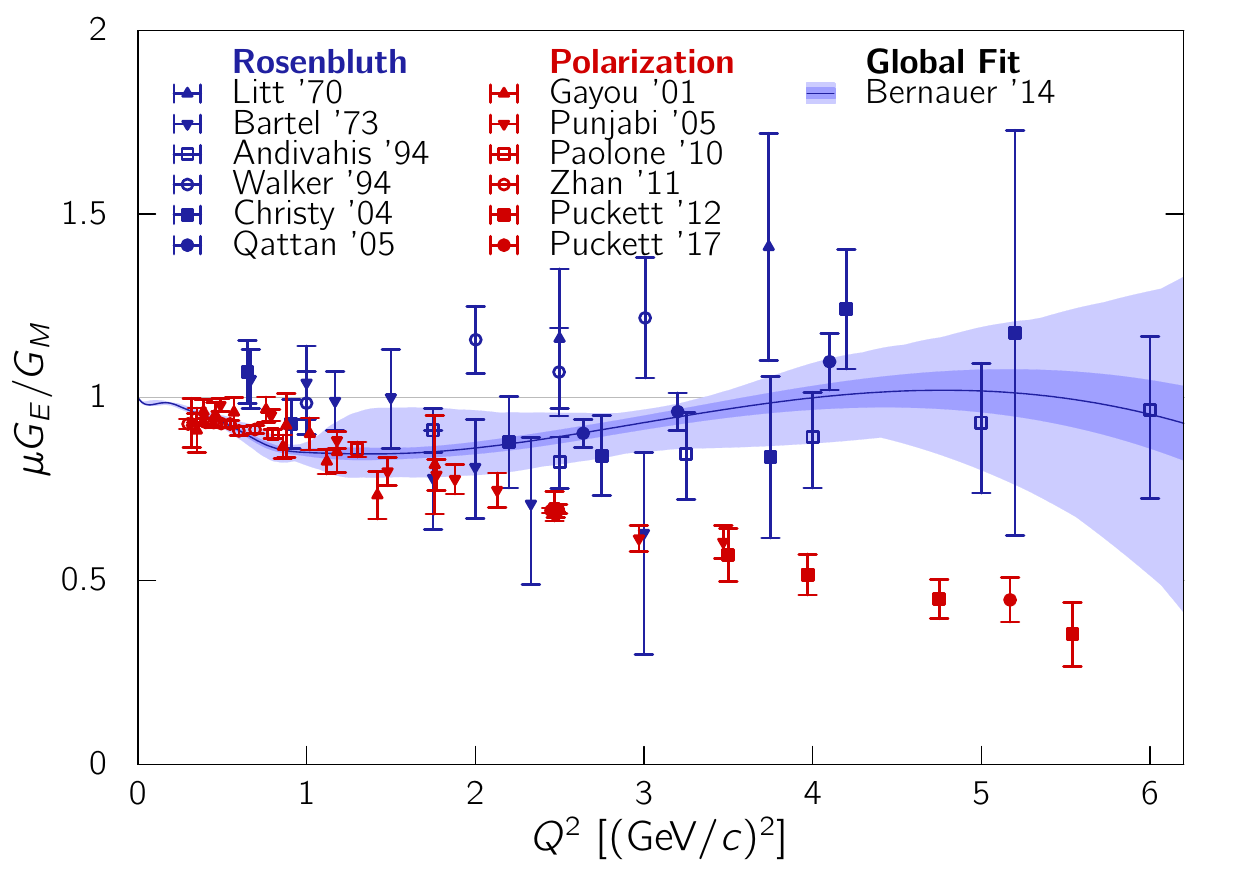}
    \caption{A representative sample of the world data on the proton's form factor ratio, $\mu_p G_E/G_M$ shown as a function of squared four-momentum transfer, $Q^2$. Rosenbluth separations of unpolarized cross sections are shown in blue \cite{Litt:1969my,Bartel:1973rf,Andivahis:1994rq,Walker:1993vj,Christy:2004rc,Qattan:2004ht}. Polarized measurements are shown in red \cite{Gayou:2001qt,Punjabi:2005wq,Paolone:2010qc,Zhan:2011ji,Puckett:2011xg,Puckett:2017flj}. A global fit to unpolarized cross sections \cite{Bernauer:2013tpr} is shown, along with statistical and systematic uncertainties, by a blue curve with light blue bands.}
    \label{fig:ff_ratio}
\end{figure}

TPE became a serious concern for high-precision determinations of the proton's elastic form factors with the advent of the technique of polarization transfer, in the early 2000s. Measurements of polarization transfer in elastic electron-proton scattering at JLab~\cite{Jones:1999rz,Gayou:2001qd,Puckett:2010ac,Gayou:2001qt,Punjabi:2005wq,Paolone:2010qc,Zhan:2011ji,Puckett:2011xg,Puckett:2017flj,Hu:2006fy,Jones:2006kf,MacLachlan:2006vw,Ron:2011rd} and elsewhere~\cite{Milbrath:1997de,Pospischil:2001pp,Crawford:2006rz}  produced surprising results: the proton's form factor ratio, $\mu_p G_E/G_M$, falls steadily with $Q^2$.  
This trend is contrary to decades-worth of observations made using Rosenbluth separations of unpolarized cross section data~\cite{Litt:1969my,Bartel:1973rf,Andivahis:1994rq,Walker:1993vj,Christy:2004rc,Qattan:2004ht,Janssens:1965kd,Berger:1971kr}, as shown in Fig.~\ref{fig:ff_ratio}.
While the cause of this discrepancy has not been definitively determined, the leading hypothesis is that the effects of hard two-photon exchange are responsible \cite{Guichon:2003qm,Blunden:2003sp}.
Two-photon exchange cannot be calculated in a completely model-independent way and is not fully accounted for in standard approaches to radiative corrections ({\it e.g.}, Refs.~\cite{Mo:1968cg,Maximon:2000hm}; see also Ref.~\cite{Liu:2020rvc} for an alternative approach). It is possible that the two methods of extracting the proton's form factor ratio are susceptible in different ways to this effect, producing the apparent discrepancy.

\begin{figure}[t!]
    \centering
    \includegraphics[width=0.99\columnwidth]{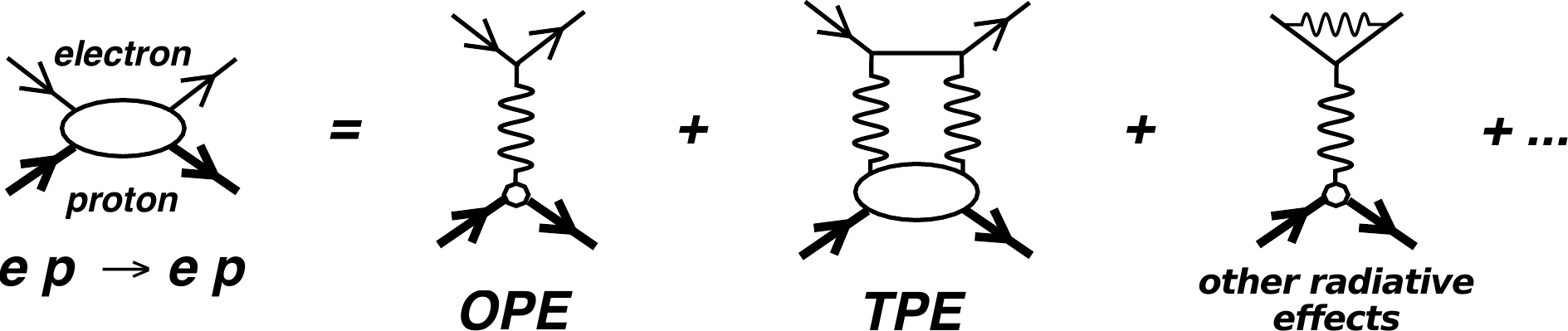}
    \caption{Feynman diagram series for elastic electron-proton scattering. The two-photon exchange amplitude contributes at the same order as several other radiative processes.}
    \label{fig:Fey_ep}
\end{figure}

Two-photon exchange is one of the sub-leading contributions to the elastic scattering amplitude, as shown in Fig.~\ref{fig:Fey_ep}, and is one of several radiative processes at the same order in the fine structure constant, $\alpha$. TPE affects the cross section at order $\alpha^3$, as an interference term between TPE and the leading OPE amplitude. Electron-scattering experiments typically report cross sections that are corrected back to the level of one-photon exchange using a radiative corrections prescription that also depends on the experiment's capabilities for resolving energy lost to soft bremsstrahlung emission. Due to the difficulties in calculating the TPE amplitude, standard prescriptions only treat TPE in the so-called ``soft limit'', in which one of the exchanged photons carries negligible 4-momentum. In this way, TPE is only partially treated; any residual effect beyond the soft-limit is termed hard TPE. Until the emergence of the proton form factor discrepancy, the effects of hard TPE were assumed to be negligibly small for almost all relevant purposes. 

The challenge in calculating hard TPE lies in fact that the diagram has an off-shell hadronic propagator. 
TPE belongs to a larger class of hadronic box dia\-grams -- including $\gamma Z$ exchange, relevant for parity-vio\-lating electron scattering \cite{Blunden:2012ty}, $\gamma W^\pm$ exchange, relevant for $\beta$-decay \cite{Marciano:2005ec} -- which can only be calculated with some degree of model dependence. 

Broadly speaking there are two theoretical approa\-ches: hadronic methods and partonic methods. In the former, the hadronic propagator is represented as a sum of contributions from all hadronic states, i.e., the nucleon, $\Delta$, $N^*$ resonances, {\it etc.}, with +1 charge and allowed spin and parity. The sum is truncated to a finite number of considered states. This approach was first employed by Blunden {\it et al.}~\cite{Blunden:2003sp}, and has since been used in numerous other calculations \cite{Blunden:2005ew,Kondratyuk:2005kk,Kondratyuk:2007hc}.
More recently, it has been further improved by using dispersion relations~\cite{Gorchtein:2006mq,Borisyuk:2006fh,Borisyuk:2012he,Borisyuk:2013hja,Tomalak:2014sva,Tomalak:2016vbf,Tomalak:2017shs,Blunden:2017nby,Ahmed:2020uso} to eliminate unphysical divergences that arise in the forward limit. Hadronic calculations suggest that TPE has a percent-level effect on the elastic cross section, and that the magnitude of the effect increases at backward angles, which may be sufficient to resolve the form factor discrepancy~\cite{Arrington:2007ux}. Hadronic calculations are expected to be valid for smaller momentum transfers, approximately $Q^2 < 3$~(GeV$/c)^2$.

\begin{figure}[t!]
    \centering
    \includegraphics[width=0.99\columnwidth]{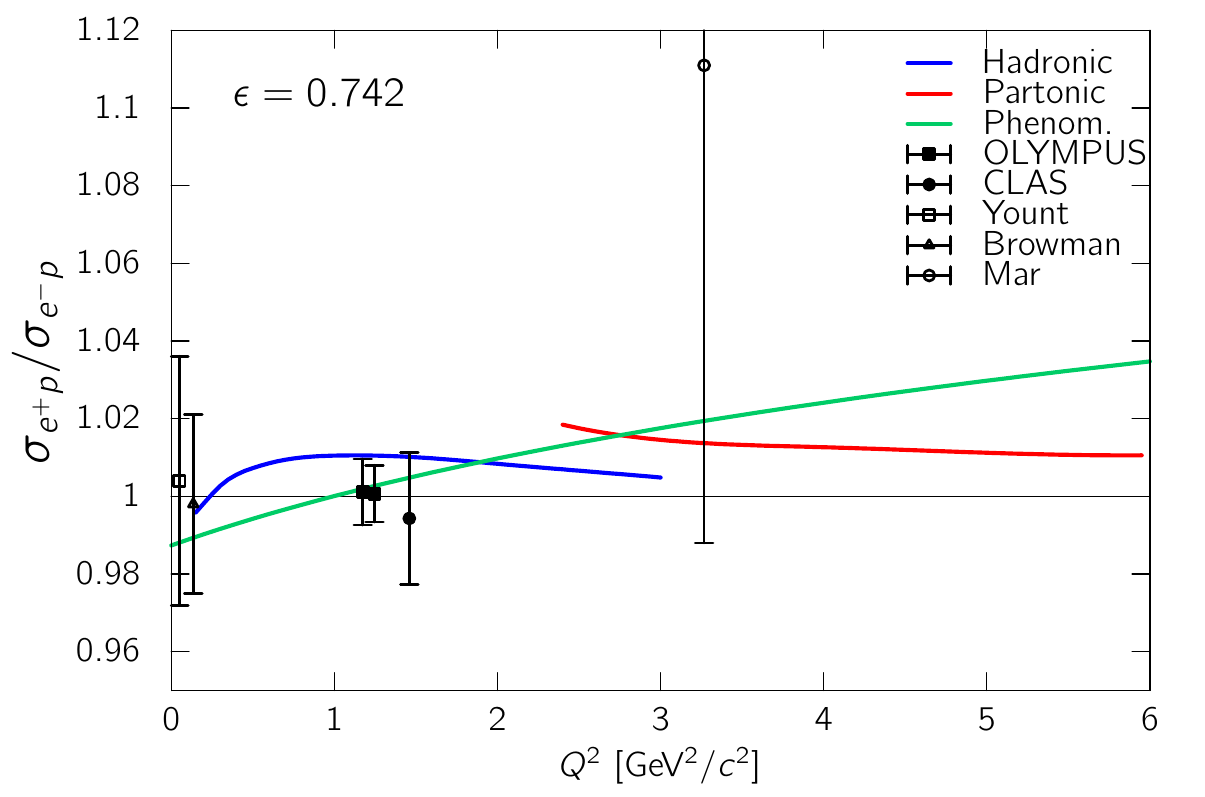}
    \caption{The positron-proton to electron-proton elastic scattering cross section ratio predicted by examples of three approaches to calculating hard TPE: a hadronic calculation (Blunden and Melnitchouk $N+\Delta$ \cite{Blunden:2017nby}) in blue, 
    a partonic calculation (Kivel and Vanderhaeghen, BLW model \cite{Kivel:2009eg}) in red,
    and a phenomenological extraction from the magnitude of the form factor discrepancy (Bernauer {\it et al.}\ \cite{Bernauer:2013tpr}) in green. The calculations are for fixed $\epsilon=0.742$, and assume the Mo and Tsai~\cite{Mo:1968cg} convention for the definition of soft TPE. Also shown are available data for $0.722 < \epsilon < 0.762$ from CLAS \cite{Rimal:2016toz}, OLYMPUS \cite{Henderson:2016dea} and measurements from the 1960s \cite{YountPine,Browman:1965zz,Mar:1968qd}.}
    \label{fig:tpe_pred}
\end{figure}

By contrast, partonic calculations of TPE should be increasingly valid in the limit of large momentum transfer. Partonic calculations model the interactions of the exchanged photons with individual quarks, whose distributions within the proton are described by GPDs ({\it e.g.} in Refs.~\cite{Chen:2004tw,Afanasev:2005mp}) or distribution amplitudes ({\it e.g.}, in Refs.~\cite{Borisyuk:2008db,Kivel:2009eg}). Such approaches must assume factorization between the hard and soft parts of the amplitude and must further model the distribution of quarks within the proton. Depending on the assumptions ma\-de, there can be a wide spread in predictions, as shown in Fig.~\ref{fig:tpe_pred} for examples of hadronic~\cite{Blunden:2017nby} and partonic~\cite{Kivel:2009eg} calculations, and a phenomenological estimate based on the size of the form factor discrepancy~\cite{Bernauer:2013tpr}.

While TPE poses significant challenges for theory, it can be determined through a number of experimental observables. Though positron-scattering is not the only way to experimentally constrain hard two-photon exchange, it is one of the best. Since the interference term between one- and two-photon exchange changes sign between electron-scattering and positron scattering, TPE induces asymmetries in many observables when measured with electrons versus positrons. In fact, three recent experiments were conducted to measure the ratio of the unpolarized positron-proton to electron-proton elastic scattering cross sections, with the goal of determining if TPE is the cause of the proton form factor discrepancy~\cite{Rachek:2014fam,Adikaram:2014ykv,Rimal:2016toz,Henderson:2016dea}. The results, while showing modest indications of hard TPE, were far from conclusive because of their limitation to low-$Q^2$ kinematics  ($Q^2<2$~(GeV$/c$)$^2$) where the form factor discrepancy is small. More decisive measurements at higher $Q^2$ and with larger beam energies are needed. The regime between $3<Q^2<5$~(GeV$/c$)$^2$ is particularly interesting because not only is the form factor discrepancy large, but it also sits between the regions where hadronic and partonic calculations are expected to work best. 

Quantifying the amount of hard TPE is important for improving our understanding of proton structure, but also for improving radiative corrections relevant to several other problems in precision electroweak physics. Until TPE can be decisively quantified over a wide kinematic range, it remains an obstacle to refining our knowledge about proton structure, both for the push to high $Q^2$, the focus of the new JLab SBS program, and at low $Q^2$ where significant uncertainty remains about the proton radius. Measurements of TPE also provide valuable constraints on model-dependent theoretical calculations of the $\gamma Z$-box corrections in parity-violating electron scattering, as well as the $\gamma W^\pm$-box, relevant for radiative corrections to $\beta$-decay lifetimes. 

Currently, among the facilities around the world that can produce positron beams, none possess both an accelerator of the energy of CEBAF and detector systems in the same league as those  operating in and planned for the JLab experimental Halls. This deficit renders a number of highly impactful potential measurements out of reach for now. A high-quality positron beam in CEBAF would permit a diverse and exciting program of measurements of two-photon exchange that would provide crucial experimental constraints, help solidify our understanding of nucleons structure, and even help test the limits of the standard model. 

This Topical Issue presents a number of experimental concepts for measurement of TPE via several different observables. Three concepts employ the most traditional approach: comparing the unpolarized elastic positron-proton scattering cross section to that of elec\-tron-proton scattering. The most comprehensive measurement could be performed with the CLAS12 detector~\cite{Burkert:2020akg} in Hall~B, where the enormous acceptance would provide unparalleled kinematic reach~\cite{Bernauer:2021vbn}, and where the typical beam currents match what the proposed positron source could provide. This could be complemented by a rapid two-week measurement, focusing on low-$\epsilon$ kinematics, in Hall~A~\cite{Cline:2021hkr}, where the planned Super BigBite Spectrometer would allow higher luminosity running. The spectrometers in Hall~C would be well-suited for performing a so-called super-Rosenbluth measurement with positrons~\cite{Arrington:2021kdp}, in which an L/T separation is performed from cross sections in which only the recoiling proton is detected. The results of a posi\-tron super-Rosenbluth measurement could be directly compared to those of a previous measurement in Hall~A, taken with electrons~\cite{Qattan:2004ht}.

Positrons would be valuable for constraining TPE through observables different from unpolarized elastic cross sections. Polarization transfer, while expected to be more robust to the effects of hard TPE, is sensitive to a different combination of generalized form factors, and a measurement with both electrons and positrons provides new constraints. A 90-day measurement~\cite{Puckett:2021gya}, at $Q^2=2.6$ and 3.4~(GeV$/c$)$^2$, would be possible in Hall~A~\cite{Alcorn:2004sb}, using Super BigBite in a similar configuration to the upcoming GEp-V experiment~\cite{GEP5_PAC47}. Super BigBite would also be useful for a measurement of the target-normal single-spin asymmetry in positron-proton scattering~\cite{Grauvogel:2021btg}. Transverse single-spin asymmetries are zero in the limit of one-photon exchange, and a nonzero asymmetry measurement can either be caused by an imaginary component in the TPE amplitude, or some unknown T-violating process. A measurement with electrons and positrons can distinguish between the two. 

In addition to high-$Q^2$ electron scattering, TPE at low $Q^2$ is a topic of special interest by itself~\cite{Talukdar:2019dko}, and has received extra attention due to its possible effects on the extraction of the proton radius~\cite{Carlson:2015jba}. The proton's charge radius, defined as the slope of the charge form factor at $Q^2=0$~(GeV$/c$)$^2$~\cite{Miller:2018ybm}, does not depend on the probe; any difference in the apparent size of the proton is an indication of higher order effects or analysis differences not being properly taken into  account~\cite{Guichon:2003qm,Blunden:2003sp,Higinbotham:2019jzd}.
The MUSE experiment~\cite{Gilman:2013eiv,Gilman:2017hdr}, which has begun running at the Paul Scherrer Institute, investigates lepton universality in electron and muon elastic scattering on the proton at low $Q^2$. Using the new Prad-II setup~\cite{Gasparian:2020hog}, electron and positron scattering at low $Q^2$ can be studied with high precision on protons as well as on  deuterium~\cite{Hague:2021xcc}, using gaseous targets as designed with the novel Prad-II target~\cite{Pierce:2021vkh}.

Lastly, measurements of TPE in elastic lepton-nuc\-leus scattering~\cite{Borisyuk:2019gym} would be useful for helping to constrain nuclear models used for calculations of $\gamma W^\pm$ box diagrams, constituting important radiative corrections in $\beta$-decay. The $\beta$-decay widths for a number of super-allowed transitions are important inputs for tests of the unitarity of the first row of the Cabibbo–Kobayashi– Maskawa matrix. Measurements of TPE via the unpolarized $e^+ A/e^- A$ cross section ratio on a number of specific isotopes can help improve the radiative corrections necessary to searching for new physics in the quark sector. A key to this measurement is the ability to resolve the events in which the nucleus remains in the ground state, but resolution of the spectrometers in Halls~A and C are more than sufficient, especially since the rates would be low enough to permit the use of drift chambers for tracking. A 25-day measurement would be sufficient to cover six different nuclei in three different kinematics to 1\% statistical precision~\cite{Kutz:2021scv}.

Two-photon exchange is important to measure not least of all to consolidate our understanding of nucleon form factors, but also because it deals with many open problems related to radiative corrections in parity violation and $\beta$-decay. For the time being, a positron beam at CEBAF would be the only feasible avenue for pursuing the broad TPE program described in this issue.

%
%

\section{Nucleon \& nuclear tomography}

Quantum chromodynamics (QCD) has been established as the theory that describes the interaction between the quarks and the gluons, the fundamental building blocks of hadronic matter. However, QCD cannot provide a method for the analytical derivation of the fundamental nucleon  properties and eventually explain the mechanism of 
their generation from the elementary partonic degrees of freedom. The primary tools for the inspection of the internal structure of the nucleon are partonic distribution functions that can be accessed in high-energy scattering processes. In particular, GPDs are nowadays the object of an intense research efforts. A proper determination of GPDs would be crucial to shed light on how QCD works. They can provide  a tomographic image of the nucleon and atomic nuclei~\cite{Burkardt:2000za,Diehl:2002he}, by correlating the longitudinal momentum and the transverse spatial position of the partons inside the nucleon, and give access to the contribution of the orbital angular momentum of quarks to the nucleon spin~\cite{Ji:1996ek}. 

The GPDs of nucleons and nuclei are accessed in the measurement of the exclusive lepto-production of either a photon ($eN\to eN\gamma$, or DVCS, and $eN\to eN\gamma^{\star} \to eNl^+l^-$, or DDVCS) or a meson ($eN\to eN m$, or DVMP). The factorization theorems establish that these scattering amplitudes are dominated by terms involving the convolution of a hard scattering kernel with the nucleon GPDs if the invariant momentum transfer squared $Q^2$ and the squared hadronic center-of-mass energy are sufficiently large~\cite{Collins:1996fb,Collins:1998be}. At leading order and leading twist, considering only the quark sector and quark-helicity conserving quantities, there are 4 GPDs for each quark flavor  $(H^q,E^q,\widetilde{H}^q,\widetilde{E}^q)$, and each depends on three variables: the invariant momentum transfer $t$ to the nucleon, the average longitudinal momentum fraction $x$ carried by the active parton, and the scaling variable $\xi$ representing the parton skewness, as well as the QCD evolution scale $Q^2$ (omitted for simplicity of notation). \newline 
At $\xi$=$0$, for which $t$=$-\mathbf{\Delta}^2_\perp$, an impact parameter ($\mathbf{b}$)  version of GPDs can be derived through the Fourier integral
\begin{equation}
\rho^q_F(x,\mathbf{b}_{\perp}) = \int \frac{d^2 \mathbf{\Delta}_\perp}{(2\pi)^2} \, e^{i \mathbf{b}_{\perp}\cdot\mathbf{\Delta}_{\perp}} \, F_+^q (x,0,-\Delta^2_\perp) 
\end{equation}
where $F_+(x,0,t)$ is the $0$-skewness singlet GPD combination ($F_+^q \equiv \{H^q_+, E^q_+, \widetilde{H}^q_+, \widetilde{E}^q_+ \}$) for the quark flavor $q$, defined as
\begin{equation}
F_+^q(x,0,t) = F^q(x,0,t) \mp F^q(-x,0,t) \, ,
\end{equation}
with $0 \le x \le 1$; the upper sign holds for vector GPDs $(H^q,E^q)$ and the lower for axial vector GPDs $(\widetilde{H}^q,\widetilde{E}^q)$. For instance, $\rho_H^q(x,\mathbf{b}_{\perp})$ can be interpreted as the density of quarks of flavor $q$ with longitudinal momentum fraction $x$ at a  transverse position $\mathbf{b}_{\perp}$ from the  nucleon  center-of-mass~\cite{Burkardt:2000za}, which founds the basis ground for the tomography of hadrons. \newline
The skewness dependency of GPDs contains unique information about the nuclear dynamics. It leads to the polynomiality property of the GPDs that allows one to express the first Mellin moment of the GPD $H$ as 
\begin{equation}
\int_{-1}^{1} dx \, x \, \sum_q H^q(x,\xi,t) = M_2(t) + \frac{4}{5} \xi^2 d_1(t) \, ,
\end{equation}
where $d_1(t)$ is the first coefficient of the Gegenbauer expansion of the $D$-term, and  only the forward limit ($t \to 0$) of $M_2(t)$ is known, from the momentum distribution of quarks and anti-quarks at the QCD scale  $Q^2$~\cite{Goeke:2001tz}. $d_1(t)$ reflects the internal dynamics of a hadron through the distribution of forces~\cite{Polyakov:2018zvc}. It is the gravitational form factor of the energy-momentum tensor (EMT) which encodes the distribution of pressure and shear forces inside hadrons~\cite{Ji:1996ek}. While it is hopeless to expect direct observation of the interaction of a graviton with a hadron, GPDs offer a unique indirect way to access these properties. The relation between the GPDs and the EMT of the nucleon also offers the ability to resolve the long-standing puzzle of the decomposition of the nucleon spin. This is expressed by the Ji's 
sum rule~\cite{Ji:1996ek}
\begin{equation}
\lim_{t \to 0} \int^1_{-1} dx \, x \, \left[ H^q(x,\xi,t) + E^q(x,\xi,t) \right] = J^q,
\end{equation}
which links the forward limit of the sum of the second moments of the GPDs $H^q$ and $E^q$ to the total angular momentum carried by the quarks inside the nucleon. \newline
Accessing nucleon tomography or the QCD dynamics of the nucleon asks for the mapping of the $x$-, $\xi$-, and $t$-dependences of the GPDs over the full physics phase-space, an evidently ambitious and demanding  experimental program.

\begin{figure}[t!]
    \centering
    \includegraphics[width=0.99\columnwidth]{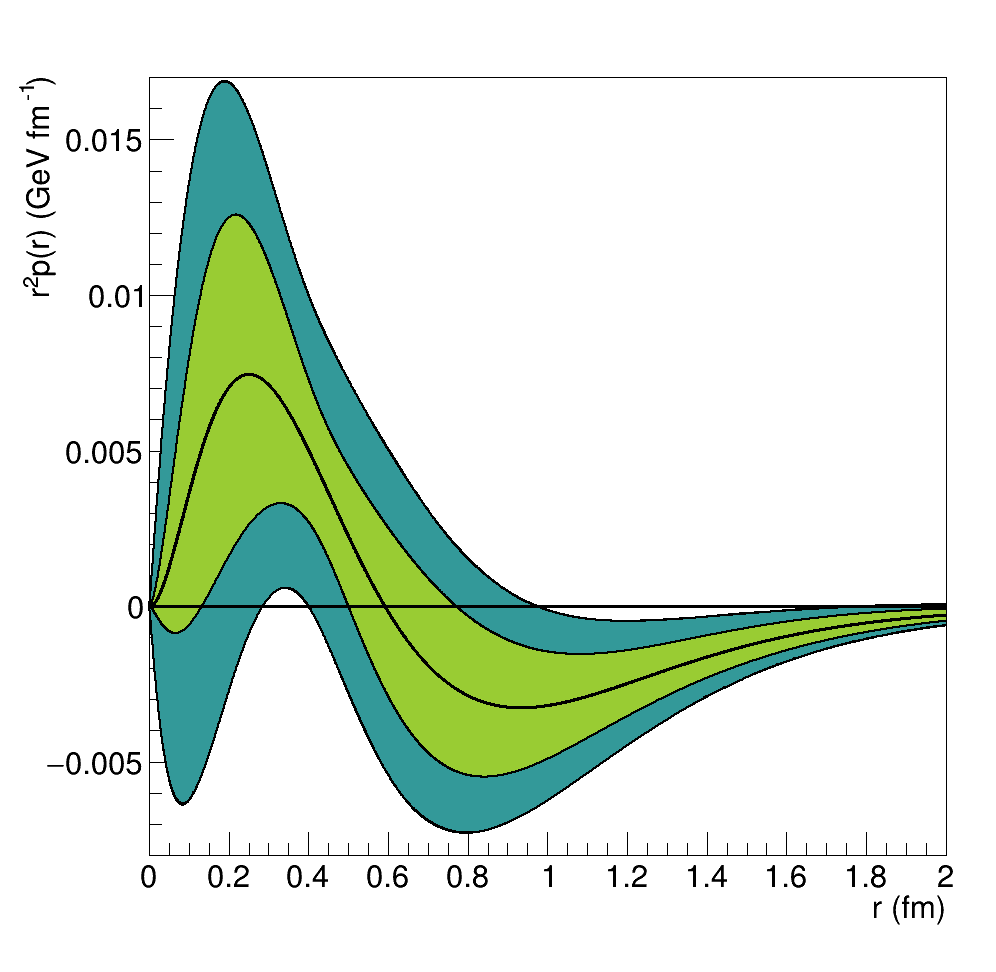}
    \includegraphics[width=0.99\columnwidth]{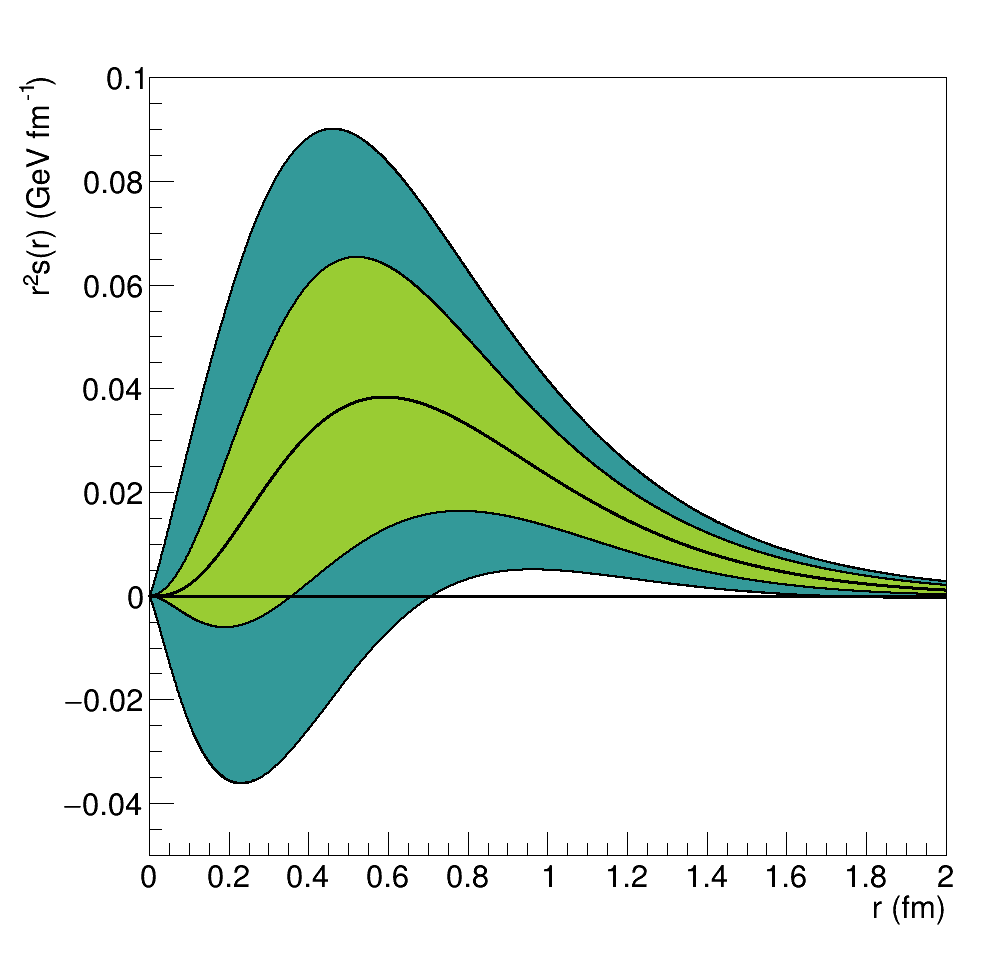}
    \caption{Radial distribution of the pressure $r^2 p(r)$ (top) and shear forces $s(r)$ (bottom) resulting from the interactions of the quarks in the proton~\cite{Burkert:2018bqq,Burkert:2021ith}. The middle lines corresponds to the information extracted from the $D$-term fitted to DVCS CLAS data at 6~GeV. The  bands represent the range of uncertainties without (outer band) and with (inner band) CLAS data.}
    \label{fig:forces}
\end{figure}
The GPDs do not enter directly in the $DVCS$ amplitude, but only as combinations of integrals over the average light-cone momentum fraction $x$. The remaining variables are purely kinematic, in that they are measured event-by-event in the scattering process. These integrals are referred to as Compton form factors (CFFs) ${\mathcal F}$ (with ${\mathcal F} \equiv \{ {\mathcal H}, {\mathcal E}, \widetilde{\mathcal H}, \widetilde{\mathcal E} \}$) defined at the leading order (LO) approximation as  
\begin{eqnarray}
{\mathcal F}(\xi,t) & = & {\mathcal P} \int_{0}^{1} dx \left[ \frac{1}{x-\xi} \pm \frac{1}{x+\xi} \right] F_+(x,\xi,t) \nonumber \\
& & \,\,\,\,\,\, - \, i \pi \, F_+(\xi,\xi,t), \label{eq:CFF}  
\end{eqnarray}
where ${\mathcal P}$ denotes the Cauchy's principal value integral, and 
\begin{equation}
F_+(x,\xi,t) = \sum_{q} \left(\frac{e_q}{e}\right)^2 {\left[ F^q(x,\xi,t) \mp F^q(-x,\xi,t) \right] } \, .
\end{equation} 
Though the GPDs are purely real functions, the CFFs are complex-valued. Analytical properties of the $DVCS$ amplitude at the leading order approximation link the real and imaginary parts of the CFFs through the dispersion relation~\cite{Anikin:2007yh,Diehl:2007jb,Polyakov:2007rv}
\begin{eqnarray}
\Re {\rm e} \left[ {\mathcal F}(\xi,t) \right] & \stackrel{\rm LO}{=} & \Delta_{\mathcal F}(t) \\ 
& & \hspace*{-45pt} + \frac{1}{\pi}{\mathcal P}\int_{0}^{1} dx  \left
( \frac{1}{\xi-x}-\frac{1}{\xi+x}\right) \Im{\rm m} [{\mathcal F}(x,t)], \nonumber
\end{eqnarray}
where $\Delta_{\mathcal F}(t)$ is a $t$-dependent subtraction constant related to the $D$-term. Thus, the independent knowledge of the real and imaginary parts of the CFFs allows us to access the nucleon dynamics. This feature was remarkably  developed in recent  works~\cite{Burkert:2018bqq,Burkert:2021ith} determining the radial distribution of pressure and shear forces in the proton from existing DVCS data (Fig.~\ref{fig:forces}). Considering the present status of experimental knowledge of GPDs and the resulting lack of constraint with respect to the hypotheses formulated to extract the $D$-term, the precise shape of the derived distribution should be taken with caution~\cite{Kumericki:2019ddg,Dutrieux:2021nlz}. However, these curves clearly demonstrate the physics potential of DVCS data with respect to the investigation of QCD dynamics, and advocate for the unambiguous measurements of the real and imaginary parts of the CFFs.

Given the complexity of the GPDs and their complicated link to experimental observables, their measurement is a highly non-trivial task. This necessitates a long-term experimental program comprising the measurement of different DVCS observables (to single out the contribution of each of the 4 GPDs), on the proton and on the neutron (to disentangle the quark-flavor dependence of the GPDs): cross sections, beam-, longitudinal and transverse 
target-single polarization observables, double polarization observables, and beam-charge asymmetries. Such dedicated experimental program, concentrating on a proton target, has started worldwide in these past few years. 

After the first observations of a $\sin(\phi)$ dependence for the $\vec{e}p\to ep\gamma$ reaction in low statistics beam-spin asymmetry measurements by the  HERMES~\cite{Airapetian:2001yk} and CLAS~\cite{Stepanyan:2001sm} collaborations, various high-statistics DVCS experiments were per\-for\-med. The HERA collider experiments measured  cross sections for DVCS at high $Q^2$ and low $x_B$~\cite{Aaron:2007ab,Chekanov:2008vy}. Polarized and unpolarized cross sections measured at JLab Hall A indicated, via a $Q^2$-scaling test, that the factorization and leading-twist approximations dominate the cross sections (at the $\sim 80\%$ level) already at relatively low $Q^2$ ($\sim 2$~(GeV/$c$)$^2$) in the quark valence region~\cite{MunozCamacho:2006hx}. High-statistics and wide-coverage beam-spin  asymmetries~\cite{Girod:2007aa} and cross sections~\cite{Jo:2012yt} measured in Hall B with CLAS, brought important constraints for the parameterization, in particular, of the imaginary part of the CFF of the GPD $H$. These data were expanded with results from JLab experiments at 6~GeV of longitudinally polarized target-spin asymmetries along with double-polarization observables, which provided a first look at the imaginary part of the CFF of the GPD $\widetilde{H}$~\cite{Seder:2014oaa}. Initial constraints on the $E$ GPD, crucial to the Ji spin sum rule, were obtained with DVCS measurements on the neutron~\cite{Mazouz:2007aa} and on a transversely polarized proton~\cite{Airapetian:2011uq}. These data have led to many empirical models and model-based global fits of  GPDs~\cite{Goeke:2001tz,Vanderhaeghen:1999xj,Guidal:2004nd,Kumericki:2007sa,Mezrag:2013mya,Kumericki:2016ela,Moutarde:2019tqa}.
 
The energy upgrade of the CEBAF to 12 GeV was undertaken in order to pursue the experimental study of the confinement of quarks and of the three dimensional quark-gluon structure of the nucleon with a particular focus on the study of GPDs. An extensive program is ongoing in the Halls A, B, and C, on both proton and neutron DVCS observables with polarized beam and targets, with wide acceptance (CLAS12) and with high luminosity (Halls A and C). The addition of a polarized positron beam to the CEBAF accelerator would open up the perspective of measuring new GPD-related observables, specifically beam-charge dependent asymmetries (BCAs).

\begin{figure}[t!]
    \centering
    \includegraphics[width=0.99\columnwidth]{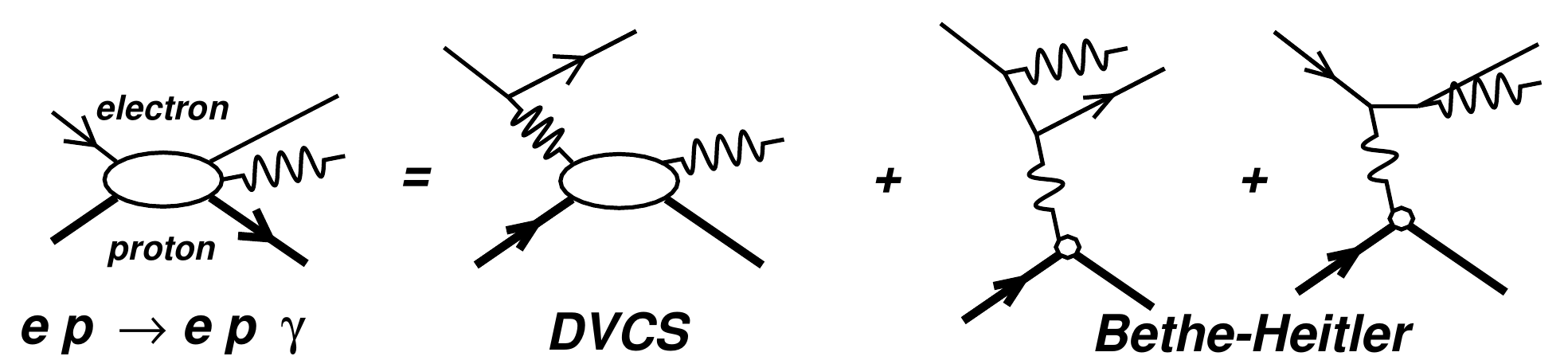}
    \caption{Lowest-order diagrams of the $eN\gamma$ process featuring the $DVCS$ and the $BH$ reaction amplitudes.}
    \label{fig:eNg}
\end{figure}
For instance, the five-fold differential cross section of the $\vec{e}N \to eN\gamma$ reaction (Fig.~\ref{fig:eNg}) - involving the interaction of a longitudinally polarized lepton beam of helicity $\lambda$ and charge $e$ with an unpolarized nucleon - may be expressed as~\cite{Diehl:2009:genova}
\begin{eqnarray}
d^5\sigma_{\lambda}^{e} & = & d^5\sigma_{BH} + d^5\sigma_{DVCS} \nonumber \\
& + &  \lambda \, d^5\widetilde{\sigma}_{DVCS} - e \left[ d^5\sigma_{INT} + \lambda \, d^5\widetilde{\sigma}_{INT} \right] \, , \end{eqnarray}
where the $BH$ index denotes the pure Bethe-Heitler reaction amplitude (the elastic $ep$ amplitude with the detected real photon emitted by either the initial or final electron), the $DVCS$ index denotes the pure $\gamma^\ast N\to \gamma N$ ones, and the $INT$ index represents the interference amplitude between these two mechanisms; here the $d^5\sigma_i$'s are the beam-helicity independent contributions to the cross section, and the $d^5\widetilde{\sigma}_i$'s are the beam-helicity dependent ones. At small $t$, the $BH$ amplitude is accurately calculable from the electromagnetic form factors of the nucleon such that the  $d^5\sigma_{\lambda}^{e}$ cross section involves 4 unknown quantities. Comparing lepton beams of opposite helicities, the beam spin-dependent and -independent parts of the cross section can be determined. Comparing lepton beams of opposite charges, the $INT$ contributions can be separated from the $DVCS$ ones. Therefore, the combination of polarized electron and positron beams isolates the 4 unknown components of the $\vec{e}N\gamma$ cross section out-of-which GPDs are determined, and similarly for polarized targets~\cite{Voutier:2014kea}. \newline
Beam and target single-spin dependent cross sec\-tions are proportional to the imaginary part of the interference amplitude. Thus the difference of polarized electrons or polarized positrons DVCS cross sections gives nearly direct access to the imaginary part of the CFFs, which are in turn equal (at leading order) to the GPDs on the {\it diagonal} $x$=$\pm \xi$. In addition, the DDVCS process which involves a final time-like virtual photon allows to access the $x \neq \pm \xi$ phase-space~\cite{PhysRevLett.90.012001,PhysRevLett.90.022001}. Beam-charge dependent observables in DVCS have the unique property to isolate the contributions from the real-part of the interference amplitude. 

While beam and target single-spin asymmetries are proportional to the imaginary part of the $DVCS$-$BH$ interference amplitude, accessing the real part is significantly more challenging. It appears in the unpolarized cross sections for which either the $BH$ contribution is dominant, or all three terms (pure $BH$, pure $DVCS$, and interference amplitudes) are comparable. The $DVCS$ and $INT$ terms can be separated in the unpolarized cross-sections by exploiting their dependencies on the incident beam energy, a generalized Rosenbluth separation. This is an experimentally elaborated procedure, and necessitates some theoretical hypothesis to extract the physics content~\cite{Defurne:2017paw,Kriesten:2020apm}. The real part also appears in double-spin asymmetries, but these can receive  significant direct contribution from the BH process itself, and are also experimentally challenging. Unpolarized BCAs are directly proportional to the real part of the $INT$ term, and receive no direct contribution from the BH process. As such they provide the cleanest access to this crucial observable, without the need for additional theoretical assumptions in the CFFs extraction procedure.

The present Topical Issue conjugates this feature with several experimental scenarios addressing the real part of CFFs through the direct comparison of electron and positron cross sections or BCA observables. In Hall C, the association of the High Momentum Spectrometer with the Neutral Particle Spectrometer  would enable high precision $e^+p\gamma$ cross section measurements at selected kinematics~\cite{Munoz:2021:WP}. Compared with electron beam data~\cite{munoz:2010:prop} to come within the next years, a precise determination of the real part of the CFFs $\mathcal{H}$ and $\widetilde{\mathcal{H}}$ would be achieved. Polarized and unpolarized BCA observables off the  proton~\cite{Burkert:2021rxz} would be measured using the CLAS12 spectrometer, enabling the mapping of the real part of the CFF $\mathcal{H}$ over a wide kinematical domain and probing the relative importance of higher-twist effects. Similarly, polarized and unpolarized BCA observables off the neutron~\cite{Niccolai:2021qjm} could also be measured, allowing us to extract the real part of the CFF $\mathcal{E}_n$ and $\widetilde{\mathcal{H}}_n$, ultimately leading to the quark-flavor separation of the CFFs. Complementing CLAS12 with the ALERT low energy recoil tracker~\cite{ALERT} will permit the investigation of coherent and incoherent DVCS off nuclei~\cite{Scopetta:2021:WP}, providing a novel method to look at nuclear forces and modifications of the nucleon structure through the real part of the CFFs. The addition of a muon detector to the SoLID spectrometer would enable measurements of polarized electron and positron beams DDVCS cross sections, giving a direct access to the real and imaginary parts of the CFF $\mathcal{H}(\xi',\xi,t)$ related  to the GPD out-of the diagonals $x$=$\pm\xi$~\cite{Zhao:2021zsm}. 

A program of both electron and positron scattering with CEBAF at JLab (and the future Electron Ion Collider) would have much greater impact than simply a quantitative change of GPD uncertainties. Direct access to the real part of the CFFs would be a qualitative shift for 3-D imaging of nucleons and nuclei. The measurement of DVCS with a positron beam is a key factor for the completion of the ambitious scientific program for the understanding of the 3-D structure and dynamics of hadronic matter.

%
%

\section{Tests of the Standard Model}

Our understanding of the Standard Model of particle physics reached an important milestone in 2012, brought about by the experimental observation of the Higgs boson by the ATLAS and the CMS collaborations at the LHC~\cite{Aaboud:2018zhk,Sirunyan:2018kst}. Since then, the research of both medium- and high-energy particle physics has focused on high precision tests of the Standard Model (SM) and searching for beyond-the-Standard-Model (BSM) physics. Most recently, experimental results on the $b$ quark decay~\cite{Aaij:2021vac} and the muon $g-2$ measurement~\cite{PhysRevLett.126.141801} raised challenges to lepton universality, adding fresh and exciting information to the field. 

The CEBAF has provided an essential tool in our pursuit of understanding the strong interaction and the nucleon and nuclei structure since the late 1990's. In the recent decade, studies of electroweak (EW) physics has emerged as a new direction for the JLab research program, and is complementary to high-energy experiments, adding unique information to Standard Model research worldwide.  A positron beam at JLab will open up new possibilities to test the Standard Model. In the following we focus on three specific examples: the measurement of a new set of EW  neutral current (NC) couplings ($g_{AA}^{eq}$), the investigation of charged lepton flavor violation (CLFV), and the search for BSM dark photons.

\subsection{Access to the $g^{eq}_{AA}$ electroweak couplings}

At energies much below the mass of the $Z^0$ boson (the $Z$-pole), the Lagrangian of the EW NC interaction relevant to the deep inelastic scattering (DIS) of electrons off quarks inside the nucleon is given by~\cite{10.1093/ptep/ptaa104} 
\begin{eqnarray}
L_{NC}^{eq} &=& \frac{G_F}{\sqrt{2}} \sum_q \left[ 
g_{VV}^{eq} \, \bar e\gamma^\mu e \bar q\gamma_\mu q + 
g_{AV}^{eq} \, \bar e\gamma^\mu\gamma_5 e \bar q\gamma_\mu q \right. \nonumber \\ 
&+& \left. g_{VA}^{eq} \, \bar e\gamma^\mu e \bar q\gamma_\mu \gamma_5 q + 
g_{AA}^{eq} \, \bar e\gamma^\mu \gamma_5 e \bar q\gamma_\mu \gamma_5 q \right], \label{eq:L}
\end{eqnarray}
where $G_F$ is the Fermi constant.
The $g_{VV}^{eq}$ terms are typically omitted because their chiral structure (vector-vector or $VV$) is identical to, and thus is inseparable from, electromagnetic interactions of QED. The other four-fermion couplings can be measured experimentally. 
The coupling $g_{AV}^{eq}$ was best determined in atomic parity violation experiments~\cite{Wood:1997zq,Guena:2005uj,Toh:2019iro}, while $g_{AV}^{eq}$, $g_{VA}^{eq}$ and $g_{AA}^{eq}$ can be measured in lepton scattering off a nucleon or nuclear target. Any discrepancy between their experimentally extracted and Standard Model values could point to BSM physics.

Recent parity-violating electron scattering experiments at JLab have improved the precision of the $g_{AV}^{eq}$ and $g_{VA}^{eq}$ couplings~\cite{Androic:2018kni,Wang:2014bba,Wang:2014guo}, which correspond to the axial-vector ($AV$) and the vector-axial ($VA$) chiral structures of the NC interaction between leptons and quarks, respectively. In contrast, there exist only one measurement on the axial-axial ($AA$) coupling, using the muon beams at CERN~\cite{Argento:1982tq}. Their results give $2g_{AA}^{\mu q} - g_{AA}^{\mu q} = 1.57 \pm 0.38$ which can be compared to the tree-level SM value of $1.5$. 
However, the $g_{AA}^{eq}$ couplings for electrons have never been measured directly due to a lack of high-luminosity and high-energy positron beams. The addition of positron beams to CEBAF would open up the possibility of measuring lepton-charge asymmetry between positron and electron scattering and accessing $g_{AA}^{eq}$. More specifically, the asymmetry $A^{e^+e^-}$ between unpolarized $e^+$ and $e^-$ beams DIS off a deuterium target has an electroweak contribution that is directly proportional to the combination $2g_{AA}^{eu}-g_{AA}^{ed}$ \cite{Zheng:2021hcf}. 

The extraction of $g_{AA}^{eq}$ from $A^{e^+e^-}$ faces both experimental and theoretical challenges. Experimentally, unlike parity-violation experiment where the asymmetries are taken between right- and left-handed beam electrons and helicity-correlated differences in the electron beam can be controlled to high precision using real-time feedbacks, switching between $e^+$ and $e^-$ beams will take weeks and thus measurements of $e^+$ and $e^-$ scatterings must be treated as separate experiments.  Differences in beam energy, intensity, and the detection of the scattered particles between $e^+$ and $e^-$ runs will cause sizable contributions to $A^{e^+e^-}$, though fortunately these effects have a calculable kinematic-dependence and can be separated from electroweak contributions. Theoretically, electromagnetic interaction causes an asymmetry between $e^+$ and $e^-$ scatterings at the next-to-leading order (NLO) and higher levels. The QED NLO contribution varies between a factor two to five larger than the electroweak contribution to $A^{e^+e^-}$ at the $Q^2$ values of JLab's 11~GeV beam. Therefore the higher-order contributions must be calculated precisely (to $10^{-2}$ level or better) and subtracted from data. While pure QED (and probably QCD) effects can be calculated to the required precision, contributions that arise from hadronic and non-perturbati\-ve effects are challenging to quantify. We are confident that with dedicated efforts and inputs from data, extraction of $g_{AA}^{eq}$ from the measured $A^{e^+e^-}$ data is possible and the required precision on the radiative corrections can be reached in the near future.

\subsection{Charged lepton flavor violation}

A polarized positron beam at CEBAF would also provide an opportunity to probe CLFV through a search for the process $e^+N \rightarrow \mu^+ X$~\cite{YulSon}. The discovery of neutrino oscillations provided conclusive evidence that lepton flavor is not a conserved quantity. However, lepton flavor violation in the charged lepton sector has never been observed. Even though the nonzero mass of neutrinos predicts CLFV processes such as $\mu^- \to e^-\gamma$, the predicted branching fraction Br$(\mu^- \to e^-\gamma) < 10^{-54}$ ~\cite{Baldini} is too small, and far beyond the reach of any current or future planned experiments. However, many BSM scenarios predict higher rates, within the reach of current or future experiments. In fact, BSM scenarios based on leptoquarks or R-parity violating supersymmetry allow for tree-level CLFV mechanisms.

A polarized positron beam can play an important role in the search for the CLFV process $e^+N \rightarrow \mu^+ X$.  The H1~\cite{H1:limit} and ZEUS~\cite{HERA-ZEUS} collaborations at HERA have sets limits on this CLFV process. An 11~GeV posi\-tron beam impinging on a proton target could significantly improve on the HERA limits. Due to the much smaller center of mass energy, the cross section for the CLFV DIS process will be much smaller than at HERA. However, the CEBAF facility will have an instantaneous luminosity that is larger by a factor of $\sim 10^6$ or $10^7$, allowing for an improvement over the HERA limits by up to two orders of magnitude. A polarized positron beam will also allow for independent constraints on left-handed and right-handed leptoquark states. 

 This program with high luminosity polarized posi\-trons would also complement planned CLFV studies at the future Electron-Ion Collider (EIC), where $e\to\tau$ CLFV transitions between the first and third generation leptons will be  investigated~\cite{Gonderinger_2010,Cirigliano:2021img,Boer:2011fh,AbdulKhalek:2021gbh}. For CLFV transitions between the first two lepton generations, the CEBAF positron facility is still expected to provide stronger constraints.

\subsection{Search for BSM particles}

The $e^+e^-$ annihilation process is a promising channel to search for light dark matter (LDM). LDM is a new compelling hypothesis that identifies dark matter with new sub-GeV “hidden sector” states, neutral under standard model interactions and interacting with our world through a new force mediated by a new boson:  the {\it dark photon} or $A'$.
Experiments with positron beams are particularly interesting since, for any given beam energy, there is a range of masses where the dark boson can be produced through positron resonant annihilation on atomic electrons in the target,  yielding a huge enhancement in the production rate. The availability of high energy, continuous, and high intensity positron beams at JLab would allow probing large unexplored regions in the dark photon parameter space.  

Two complementary experimental setups have been proposed~\cite{Batta:Marsi:2021}. The first makes use of a thin target to produce $A'$s through the annihilation process $e^+ e^- \rightarrow A' \gamma$. By measuring the emitted photon, the mediator of the DM-SM interaction will be identified  and its (missing) mass measured. The program proposed at JLab represents an extension of the PADME experiment. This pioneering measurement is currently taking data with the low energy positron beam available at the Laboratori Nazionali di Frascati in Italy. The higher energy positron beam available at JLab will extend the mass range by a factor of four with two orders of magnitude higher sensitivity to the DM-SM coupling constant.

The second uses a thick active target and a total absorption calorimeter to detect remnants of the light dark matter production in a missing energy experiment. Exploting the $A'$ resonant production by positron annihilation on atomic electrons, the $A'$ invisible decay will be identified by the resulting  peak in the missing energy distribution, providing a clear experimental signature for the signal. This experiment has the potentiality to  cover a wide area of the parameter space and hit the thermal target with sensitivity to confirm or exclude some of the preferred light DM scenarios. 

Although LDM models represent a particularly interesting target, the proposed experimental setups can be used more generally to search for a large range of feebly interacting particles. In particular, dark photon limits straightforwardly apply to any invisibly-decaying vector boson.

Besides the proposed program that does not rely on polarized positrons, polarization observables are expected to provide significant leverage to suppress background to identify the experimental physics signal of interest extending the reach of the above mentioned experiments. The availability of a positron beam will make JLab an ideal facility to explore the Dark Sector and BSM physics.

%
%

\section{Positron beams at JLab}

The prospect of polarized as well as unpolarized posi\-tron beams for nuclear physics experiments at CEBAF naturally raises many issues, in particular the  generation of posi\-trons and their formation into beams acceptable to the 12 GeV CEBAF accelerator.

\subsection{Polarized Electrons for Polarized Positrons}

The theoretical investigation of polarization phenomena in electromagnetic processes~\cite{Sommerfeld:1931,PhysRev.81.467.2,PhysRev.84.265}, precisely the polarization of the bremsstrahlung radiation generated by an electron beam in the vicinity of a nuclear field~\cite{Olsen:1959zz} drove the development of polarized photon beams: an unpolarized electron beam is predicted to generate a linearly polarized photon beam, while a polarized electron beam would generate a circularly polarized photon beam with polarization  directly proportional to the electron beam initial  polarization. These features were used extensively at numerous accelerator facilities, and more recently in the experimental Hall~B~\cite{Mecking:2003zu} and D \cite{Adhikari:2020cvz} of JLab to operate high energy polarized photon beams. 

\begin{figure*}[t!]
\begin{center}
\includegraphics[width=0.85\textwidth]{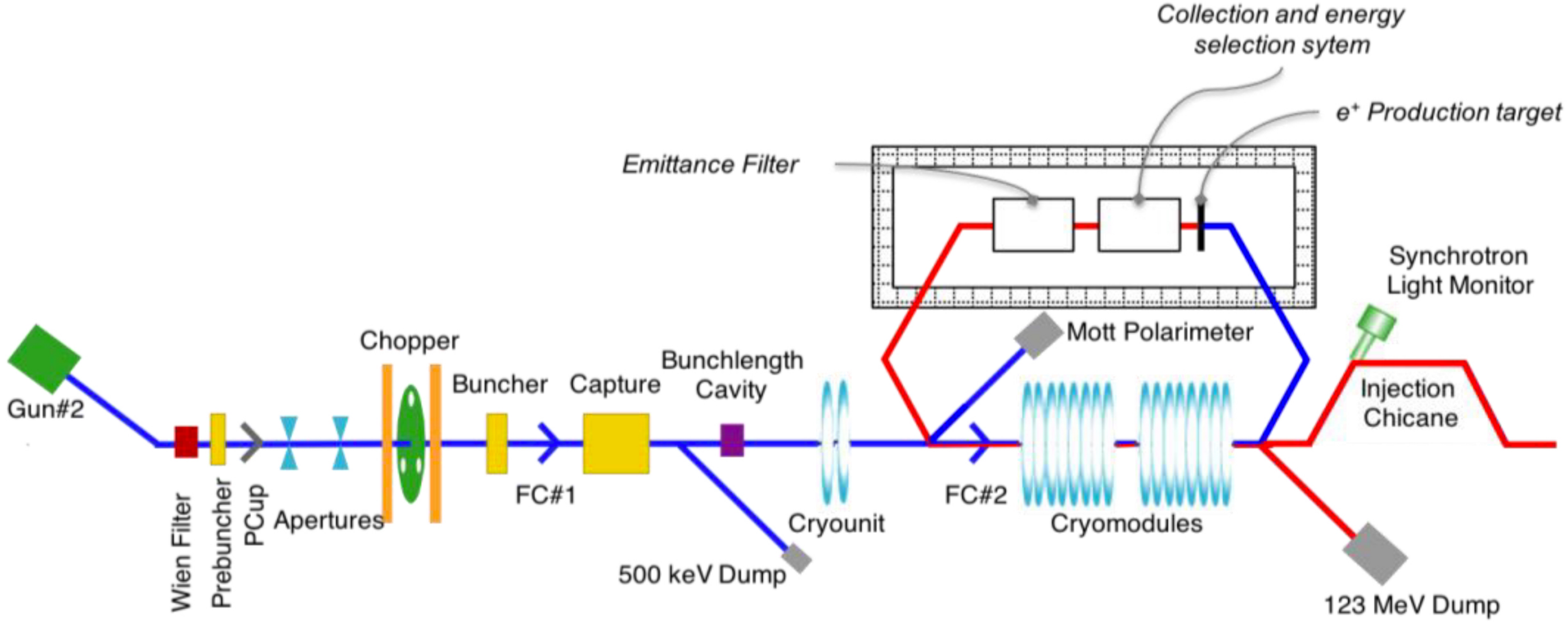}
\caption{Conceptual scheme of the integration of a positron source into CEBAF: polarized electrons (blue line) generated at the gun are accelerated up to 120~MeV/$c$ and deviated at the end of the injector into a new tunnel dedicated to positron beam production and formation; at the end of the positron source system, symbolized by its three main elements, the positron beam (red line) is deviated to enter the main accelerating section of the injector before final acceleration into CEBAF.}
\label{PEPPo@JLab}
\end{center} 
\end{figure*}
As a reciprocal process to bremsstrahlung, polarization  observables of the pair production process can be deduced from bremsstrahlung observables~\cite{Olsen:1959zz}, however paying special attention to finite lepton mass effects~\cite{Dumas:2009vy} which express differently in the bremsstralhung and pair creation processes~\cite{Kuraev:2009uy}. A circularly polarized photon beam is then predicted to create a polarized $e^+e^-$-pair whose longitudinal and transverse polarization components are both proportional to the circular polarization of the photon beam. The experimental demonstration of the circular-to-longitudinal polarization transfer has been carried out at KEK~\cite{Omori:2005eb}, SLAC \cite{Alexander:2008zza}, and JLab~\cite{Abbott:2016hyf} using completely different techniques for producing polarized photon beams. 

Following these  proof-of-principle experiments, the production of  polarized positrons at linear accelerator facilities may be separated in two categories: a first one requiring high-energy electron beams (from a few GeV to several tenths of GeV) available only at large scale facilities, and a second one accessible since a few MeV electron beam energies. The latter corresponds to the PEPPo concept~\cite{Abbott:2016hyf} which consists in the transfer of the longitudinal polarization of an electron beam to the posi\-trons produced by the bremsstrahlung polarized radiation of initial electrons interacting within a high $Z$ material. This technique can be used efficiently with a low energy ($\sim$10--100~MeV/$c$), high intensity ($\sim$mA), and high polarization (> 80\%) electron beam driver, providing a wide and cost-efficient access to polarized positron beams~\cite{Schirber:20016}.

\subsection{PEPPo @ JLab}

The PEPPo technique, which was  demonstrated~\cite{Abbott:2016hyf} at the CEBAF injector with 8.2~MeV/$c$ electrons, is the method selected for the production of polarized (and unpolarized) positron beams in support of the previously described physics program at JLab 12~GeV. PEPPo established the existence of a strong  correlation between the momentum and the polarization of the positrons: the larger the momentum, the higher the positron beam polarization, and the smaller the production rate. The quantity of interest, which  characterizes a polarized source and further enters the statistical error of the measurement of experimental signals sensitive to the beam polarization, is the Figure-of-Merit (FoM) corresponding to the product of the beam intensity with the square of the average polarization of the beam population. Based on simulations confirmed by PEPPo observations, the optimum FoM of the PEPPo technique is obtained at roughly half of the initial electron energy~\cite{Dumas:2011th}. In that respect, the essential differences between PEPPo and conventional unpolarized positron sources are the used of an initially polarized electron beam and the selection of high-momentum positron sli\-ces, that is a momentum region featuring high polarization transfer. Conversely, selecting low-momentum positrons would increase the positron beam intensity at the expense of a lower polarization. 
Given the rapid increase in the production efficiency - i.e. of positrons within a useful phase volume - with the energy of the initial electron beam, one might speculate that a very intense positron beam would benefit from the high electron beam energies available at CEBAF. This leads to the  formulation of different possible designs operating electron beam energies from 10~MeV up to 1~GeV. Cost-efficient and flexible operation between polarized and unpolarized modes favors moderate energy designs, where high intensity polarized electron sources~\cite{grames:2017Pos,Suleiman:2018obh} offer an appealing alternative to compensate for the loss in the positron production efficiency. Correspondingly, a conceptual scheme of a PEP\-Po source based on the 120~MeV/$c$ electrons (Fig.~\ref{PEPPo@JLab}) available at the end of the CEBAF injector section has been proposed~\cite{Cardman:2018svy}. It involves the construction of a new tunnel, next to the existing injector tunnel, where positrons are generated and formed into beams suitable for CEBAF injection. In this concept, it is proposed to use the same injector section to accelerate electrons towards the production energy and positrons towards CEBAF injection energies. Key apparatus of the positron source are the production target, the collection system, and the emittance filter device forming positron beams to match CEBAF admittance~\cite{Golge:2009zz}. 

\begin{figure}[t!]
\begin{center}
\includegraphics[width=0.99\columnwidth]{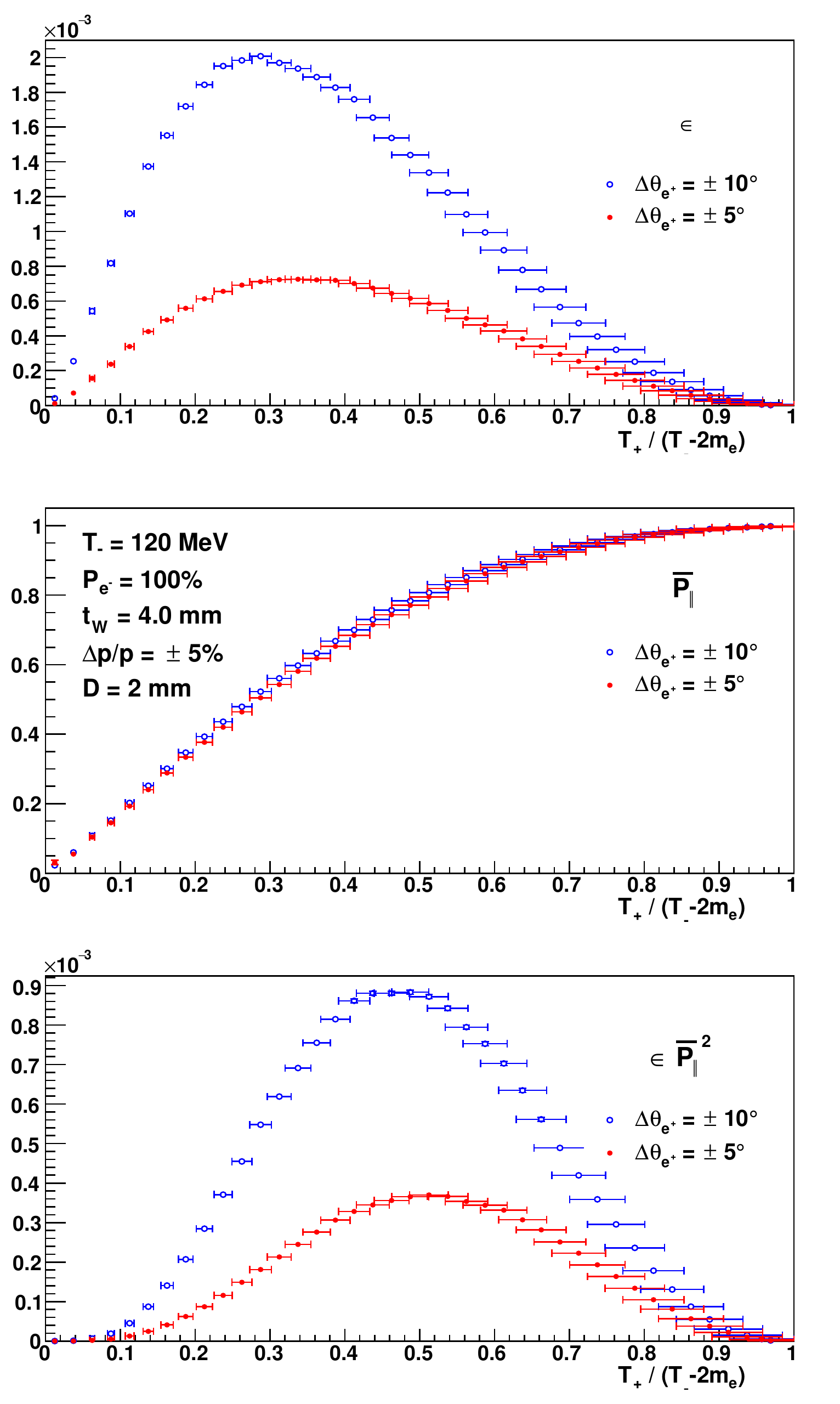}
\caption{Simulated reduced kinetic energy dependency of the positron production efficiency (top), of the average longitudinal polarization of positrons (middle), and of the FoM (bottom) for a 120~MeV  longitudinally polarized ($P_{e^-}$) electron beam impinging on a 4~mm thick tungsten target. The transverse position of positrons at the exit of the target is contained within a 2~mm diameter circular aperture. At each positron energy, the positron population within a momentum bite of $\pm$5\%, and an angular acceptance of $\pm$10$^{\circ}$ (open symbols) or $\pm$5$^{\circ}$ (closed symbols), is quantified.}
\label{PEPPo-Perf}
\end{center} 
\end{figure}
The performances of such a source, simulated with Geant4~\cite{AGOSTINELLI2003250} extended with polarization phenomena in electromagnetic  processes~\cite{Dollan:2005nj}, are shown in Fig.~\ref{PEPPo-Perf} as function of the normalized positron kinetic energy assuming a fully longitudinally polarized electron beam. They are expressed in terms of the efficiency (top pa\-nel), the average longitudinal polarization (middle pa\-nel), and the FoM (bottom panel) evaluated for a 4~mm thick tungsten target, {\it i.e.} for the optimum target thickness at 120~MeV/$c$. For each central momentum, the positron population emitted from a limited transverse area ($D$-diameter circular aperture), within a selected momentum bite $\Delta p/p$ and an angular acceptance $\Delta \theta_{e^+}$, is evaluated. This selection parameters intend to mimic the acceptance of the collection and emittance filter systems. The maximum efficiency and FoM define the source operation in unpolarized and polarized modes, respectively. The essential difference between these two modes is the energy of the positron to collect: about 1/3 of the electron beam energy for optimized efficiency, and 1/2 for optimized FoM. Angular and momentum acceptance effects strongly affects the production rate and marginally the average polarization. These parameters are driving the design of the magnetic collection system and of the RF-cavities based emittance filter device.

Even more ambitious alternative concepts may also be sketched, like starting from a positron-dedicated, high-intensity electron accelerator~\cite{Golge:2010cgi}, or implementing a PEPPo source with multi-GeV electrons. Beyond these considerations, the propagation of positrons into CEBAF is an additional concern requiring, among others, to change the polarity of arc-recirculating  magnets and to upgrade beam diagnostics. It is the purpose of the current accelerator R\&D effort to determine the  most appropriate scheme for positron beams implementation at JLab, and elaborate a conceptual design by the end of 2022. 

%
%

\section{Conclusion}

This document discussed the main physics reach of posi\-tron beams at JLab, which is further detailed in the contributions to the Topical Issue of the European Phy\-sics Journal A about {\it Positron beams and physics at Jefferson Lab (e$^+$@JLab)}. It focused on multi-photon exchange effects -- beyond the Born approximation of the electromagnetic current -- in the determination of the nucleon and nuclear electromagnetic form factors; the study of the partonic structure and dynamics of hadrons through the unambiguous determination of the real and imaginary parts of their Compton form factors; selected tests of the Standard Model looking for deviations with respect to established predictions, or the evidence of new particles characterizing possible scenarios of BSM physics; and the production of polarized and unpolarized positron beams at CEBAF.

Positron beams at JLab would open up possibilities for the decisive study of two-photon exchange phy\-sics, which is today a significant obstacle to high-preci\-sion determinations of the electromagnetic form factors. Furthermore, the immense capabilities of the existing and planned JLab detectors would offer the opportunity to quantify two-photon exchange effects in several new observables, solidifying our understanding of other hadronic box processes.

High energy and high duty-cycle positron beams at JLab would procure a tremendous qualitative shift for the study of the partonic structure of hadrons. Enabling a direct unambiguous access to the real part of Compton form factors, positron beams would provide the missing tools to establish high-precision determinations of Compton form factors and consequently generalized parton distributions. This would allow an unprecetended access to 3-D imaging and QCD dynamics of hadrons.

Positron beams would also serve the search for beyond the Standard Model physics in several channels as: the determination of the never directly measured $g^{eq}_{AA}$ electroweak  couplings via the  comparison of electron and positron deep inelastic scatterings on a deuterium target; the search for the process $e^+N \to \mu^+N$ and for left-handed and right-handed leptoquark states; and the search for dark matter particles in the $e^+e^- \to \gamma A'$ process.

This is by no means an exhaustive list of the experimental program and physics opportunities that posi\-tron beam capabilities would enable at JLab. More specific examples~\cite{Afanasev:e+TI,Bertone:e+TI,Pasquini:e+TI} are discussed in the Topical Issue and further possibilities may be foreseen, especially regarding to polarized targets where the expected posi\-tron beam intensities do not limit the experimental reach.

%

\begin{acknowledgements}

This article is part of a project that has received funding from the European Union's Horizon 2020 research and innovation program under agreement STRONG - 2020 - No~824093. It is based upon work supported by the U.S. Department of Energy, Office of Science, Office of Nuclear Physics under contract DE-AC05-06OR23177.

\end{acknowledgements}

%
%

\bibliographystyle{spphys}

\bibliography{e+ExPro}

%
%

\end{document}